\documentclass[fleqn,usenatbib]{mnras}

\usepackage{newtxtext,newtxmath}

\usepackage[T1]{fontenc}

\DeclareRobustCommand{\VAN}[3]{#2}
\let\VANthebibliography\thebibliography
\def\thebibliography{\DeclareRobustCommand{\VAN}[3]{##3}\VANthebibliography}

\usepackage{graphicx}	
\usepackage{amsmath}	

\usepackage{graphicx}
\usepackage{dcolumn}
\usepackage{bm}
\usepackage{xcolor}
\usepackage{gensymb}
\usepackage{romannum}

\title[Neutrinos from Galactic UHE sources]{
Probing LHAASO Galactic PeVatrons through gamma-ray and neutrino correspondence
}

\author[Sarmah et al.]{
Prantik Sarmah$^{1}$\thanks{prantik@iitg.ac.in}
Sovan Chakraborty,$^{1}$\thanks{sovan@iitg.ac.in}
and
Jagdish C. Joshi,$^{2}$\thanks{jagdish@aries.res.in}
\\
$^{1}$Department of Physics, Indian Institute of Technology Guwahati, Assam-781039, India\\
$^{2}$ Aryabhatta Research Institute of Observational Sciences (ARIES), Manora Peak, Nainital 263001, India\\
}

\date{}

\pubyear{2015}

\begin{document}
\label{firstpage}
\pagerange{\pageref{firstpage}--\pageref{lastpage}}
\maketitle

\begin{abstract}
Recently, Large High Altitude Air Shower Observatory (LHAASO) has detected several Galactic point sources of ultra high energy (UHE; $E_{\gamma}> 100$ TeV) gamma-rays. These gamma-rays are possibly created in leptonic or hadronic interactions of cosmic rays (CRs) of PeV energies. In the hadronic channel ($p-p$ interaction), the gamma-rays are accompanied by neutrinos. The detection of  neutrinos is therefore crucial in understanding CR acceleration in such objects. To estimate the neutrino flux, we adopt the two LHAASO sources (J2226+6057, J1908+0621) found to be spatially associated with the Supernova remnants  (SNR G106.3+2.7, SNR G40.5-0.5). For these two sources, the detected TeV-PeV gamma-ray spectra are found to be unusually hard (with spectral index $\sim$ 1.8). We develop a model of gamma-ray and neutrino emission based on the above two prototypes.   The neutrino fluxes from these two sources are found to be below the IceCube sensitivity, but are  detectable in upcoming   IceCube-Gen2 and KM3NeT experiments. We further estimate the neutrino fluxes from similar other 10 LHAASO PeVatron sources and most of them are found to be detectable in IceCube-Gen2 and KM3NeT.  Finally, we explore our model parameters, in particular the spectral power law index and estimate the future potential of the neutrino detectors to probe  CR acceleration in such Galactic sources.
\end{abstract}




\section{Introduction}
Recently, ground based Cherenkov telescope, Large High Altitude Air Shower Observatory (LHAASO) has  observed UHE (TeV-PeV) gamma-rays from 12 objects located in our Galaxy \citep{PhysRevLett.124.021102, cao2021Natur33C}. One of these sources is also detected by the Tibet AS+MD experiment and found to be associated with the supernova remnant (SNR) G106.3+2.7 \citep[][]{G1062021NatA60T}. In past, other gamma-ray experiments like  HAWC, VERITAS and HESS have detected TeV gamma-rays from some of these sources \citep{2009ApJ...703L...6A, 2018A&A...612A...1H, PhysRevLett.124.021102}.  These gamma-rays are probably connected to the knee region of the observed  cosmic ray (CR) spectrum at around 3 PeV and  are expected to be produced by the Galactic sources \citep{2013FrPhy748G}. Supernova remnants (SNRs) and pulsars  are considered to be the most favourable Galactic objects producing CRs around PeV energies \citep{2004MNRAS550B,2008ARAnA89R,2018MNRAS5237G,2019IJMPD2830022G,2021Univ324C}. The LHAASO collaboration paper \citep[][]{cao2021Natur33C},  have  listed  the SNRs and pulsars that are spatially associated with the observed gamma-rays  and are consistent with the expected potential sources of CRs \citep{montmerle1979ApJ95M, 2007ApJ131G, ahar2019NatAs61A}.  Individual supernova remnants (SNRs), i.e., IC 443 and W44, in our Galaxy have also been detected by the Fermi-Satellite in the MeV-GeV band, providing important hints of proton acceleration in SNRs \citep{pionFermi2013Sci}. Further, decade long observations of the Galactic centre region by the HESS telescope have indicated that Sagittarius A$^*$ could accelerate protons upto PeV energies \citep{hess_SgrA2016Natur76H}.

Another interesting source in our Galaxy is CRAB pulsar wind nebula (PWN), in which the radiation has been observed from radio wavelengths to PeV gamma-rays. LHAASO collaboration have also detected high energy photons from CRAB PWN upto a maximum energy of $(1.12 \pm 0.09)$ PeV \citep{2021Sci37425L}. They have interpreted the multi-wavelength radiation using a synchrotron self-Compton model with electrons having maximum energy $\approx 2.15$ PeV, hence termed as leptonic PeVatron. However, the slightly larger gamma-ray flux at the tail of the UHE spectra in CRAB PWN is considered to be of hadronic origin (CR proton acceleration) \citep{Liu2021ApJ922221L,Peng2022ApJ...926....7P}. Hence, CRAB PWN might be a source of both CR electrons and protons upto PeV energies. Indeed, the observed CR proton data by Dark Matter Particle Explorer (DAMPE) satellite indicates more than one class of Galactic CR sources \citep{An2019SciADAMPE,huang2020ApJ}.

The gamma-rays detected by LHAASO can also have the possibility of both leptonic or hadronic origin  \citep[][]{G1062021NatA60T,cao2021Natur33C}. In the leptonic channel, such gamma-rays may be produced by inverse Compton of relativistic electrons.
For the hadronic channel, the interaction of CR protons with background protons in dense gas medium can provide a dominant contribution. Gamma-rays and neutrinos are produced simultaneously in this hadronic interaction \citep{2007ApJ131G, sjtanaka2009ApJ707F}. Detection of the associated neutrinos together with the gamma-rays will probe this hadronic origin \citep{Gupta2013G_neutrinos, celli2020ApJpevatron} and will differentiate leptonic channel.

 For the hadronic channel, interaction of CR protons accelerated in SNR with molecular cloud is considered to be a prominent source of the UHE flux.  SNRs  with age ($\rm t_{\rm age} \sim 100$ yr) are favoured as accelerators of CRs upto PeV energies due to amplification of the magnetic field in the downstream shock regions \citep{2018MNRAS5237G}. The acceleration depends on the confinement time of CR protons and gas density of the acceleration zone.  Longer confinement would provide the protons with sufficient time to accelerate to higher energies. The association of UHE gamma-ray sources with old SNRs ($\rm t_{\rm age} \ge 10$ kyr) might indicate a long time confinement of CR protons, surrounding the SN shocks \citep{kargupta2022ApJ10K,Sarkar_Gupta2022_UHE}. The gas density determines  magnetic field strength and also responsible for CR energy losses due to interaction like $p-p$ collisions. For example, for a confinement time of about 10 kyr, the required source gas density is found to be about 10 $\rm cm^{-3}$  \citep[see][for details]{2022arXiv220903970S}. These accelerated protons are expected to escape the source and interact with  neighbouring molecular clouds.  Therefore, detection of neutrinos produced in these interactions is crucial to probe the origin of these PeVtrons.

 These neutrinos might be detected by current (IceCube) and future (IceCube-Gen2, KM3NeT) high energy neutrino detectors. In fact, it has been shown that IceCube and KM3NeT are sensitive to  objects like SNR G40.5-0.5 and Vela junior \citep{2015APh0M}. IceCube has detected bunch of diffuse neutrinos over a period of 7.5 years \citep[][]{2021P104b2002A}. Although, most of these  neutrinos are found to be of extra-galactic origin \citep[][]{Sarmah:2022vra,Chakraborty:2015sta,Petropoulou:2017ymv}, the possibility of few of these events originating in the galaxy can not be ruled out \citep{2013PhRvD8h1302R, 20143002N}. In addition, detection of gamma-rays by LHAASO and TibeT AS+MD makes Galactic PeVatrons strong contenders of high energy neutrino factories,  detection of neutrinos can reveal these gamma-ray associated PeVatrons.
 
 


In this work, we use the $p-p$ interaction model to estimate the resultant gamma-ray and neutrino fluxes. We built our model in the context of the two sources, SNR G106.3+2.7 and SNR G40.5-0.5. Both these sources have associated molecular cloud in close  vicinity. The gamma-rays mostly illuminated by the molecular cloud motivates for a hadronic model with protons acceleration and diffusing out of the SNR interacts with the protons in molecular cloud.  We find that neutrino flux from both these objects are below the IceCube sensitivity. We also estimate the associated neutrino flux to observed gamma-rays from the Crab nebula \citep{2021Sci37425L}, considering that the observed gamma-ray flux is completely hadronic. Even with this purely hadronic origin, the Crab neutrino flux is below IceCube limit.  We further compute the associated neutrino fluxes from the remaining LHAASO sources based on the observed gamma-ray flux. All these objects are also found to be below IceCube sensitivity.  However, many of these objects are found to be potential neutrino sources for the future neutrino detectors like IceCube-Gen2 and KM3NeT. We extended our study for the possible parameter space of our molecular cloud model; mainly the spectral index  and total energy.

This manuscript is organised as follows. In  Sec.~\ref{sec:model_of_secondaries}, we describe the model of secondary (gamma-ray and neutrino) emission in the hadronic channel and fitted the observed  gamma-ray data of the two prototype sources (SNR G106.3+2.7 and SNR G40.5-0.5). Sec.~\ref{sec:detection} is dedicated to discussion on the available parameter space and detection prospects of Galactic PeVatrons with different gamma-ray and neutrino telescopes.   Finally, we conclude this paper in Sec.~\ref{sec:conclusion}.

\begin{figure*}
    \centering
 \vbox{
 \includegraphics[width=0.48\textwidth]{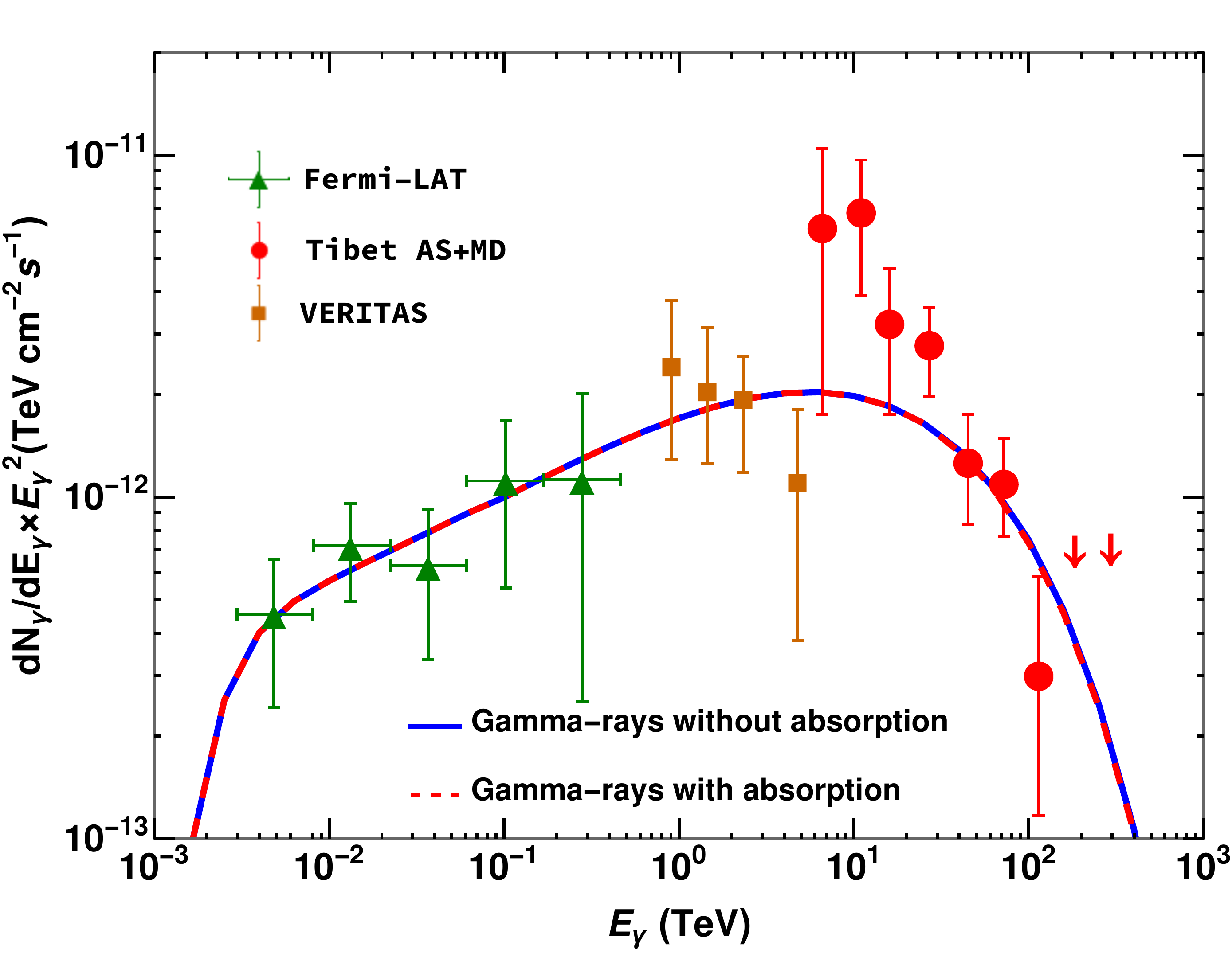}\vspace{0.5 cm}
 \includegraphics[width=0.48\textwidth]{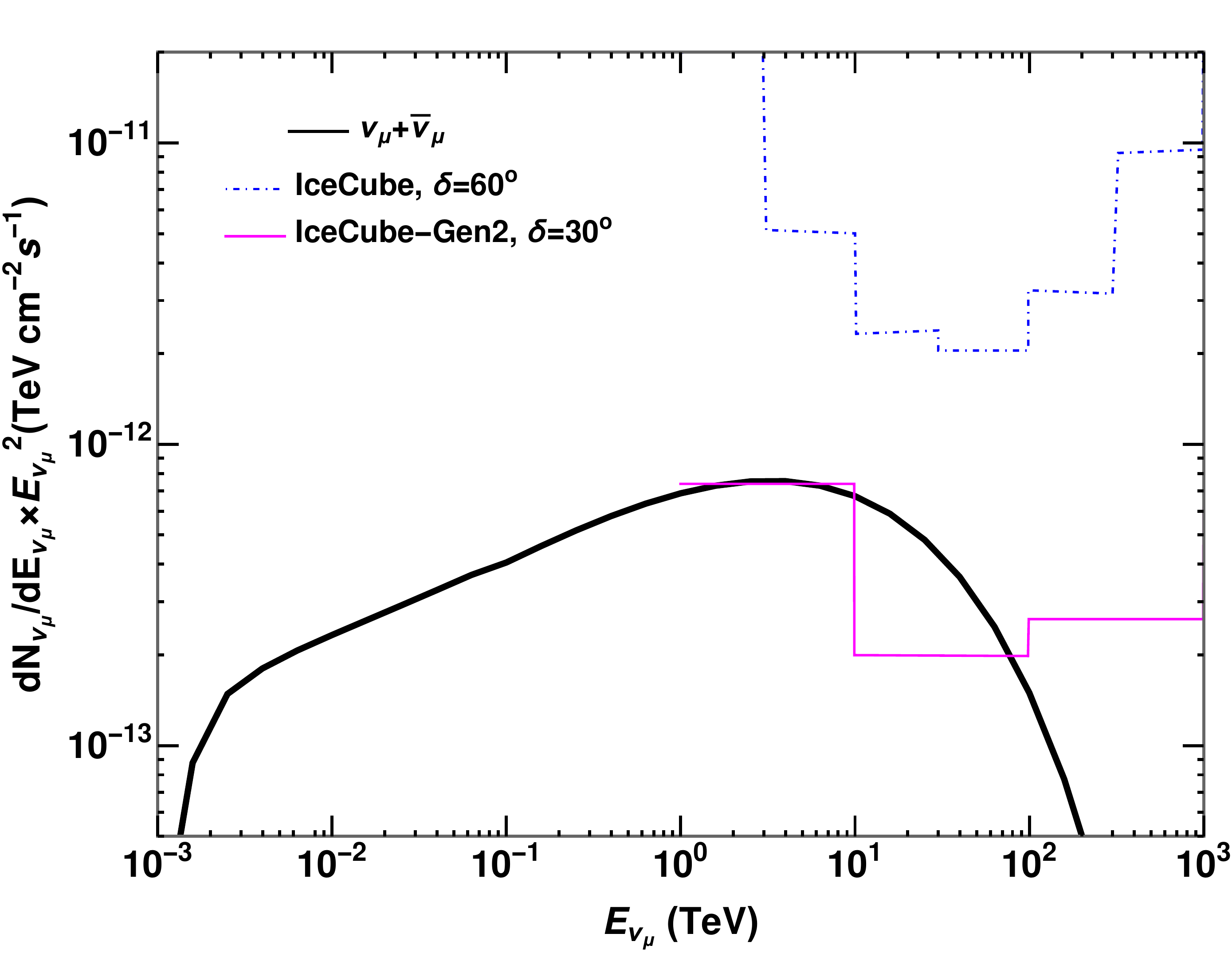}
 }
    \caption{Hadronic gamma-rays (left), and muon neutrinos (right) from SNR G106.3+2.7.   The high gamma-ray flux is fitted to the Tibet AS+MD data for a primary proton spectra with index 1.8 and the total energy of proton is $8\times10^{47}$ erg. The low energy part of the spectra is fitted to the Fermi-LAT and VERITAS data. The absorption due to pair production on ISRF is negligible that results in overlapping of the gamma-ray flux with (red-dashed) and without absorption (blue).   The distance to the source is taken to be 0.8 kpc. The correlated muon neutrino flux has been plotted (right) in addition to the detectors' sensitivities. The sensitivities of the neutrino detectors are chosen for declination, $\delta=60^{\degree}$ for IceCube and $\delta=30^{\degree}$ for IceCube-Gen2 which are closest available to the declination of this SNR. IceCube is not sensitive to this neutrino flux but IceCube-Gen2 has good sensitivity. }
    \label{fig:G106}
\end{figure*}

\begin{figure*}
    \centering
 \vbox{
 \includegraphics[width=0.48\textwidth]{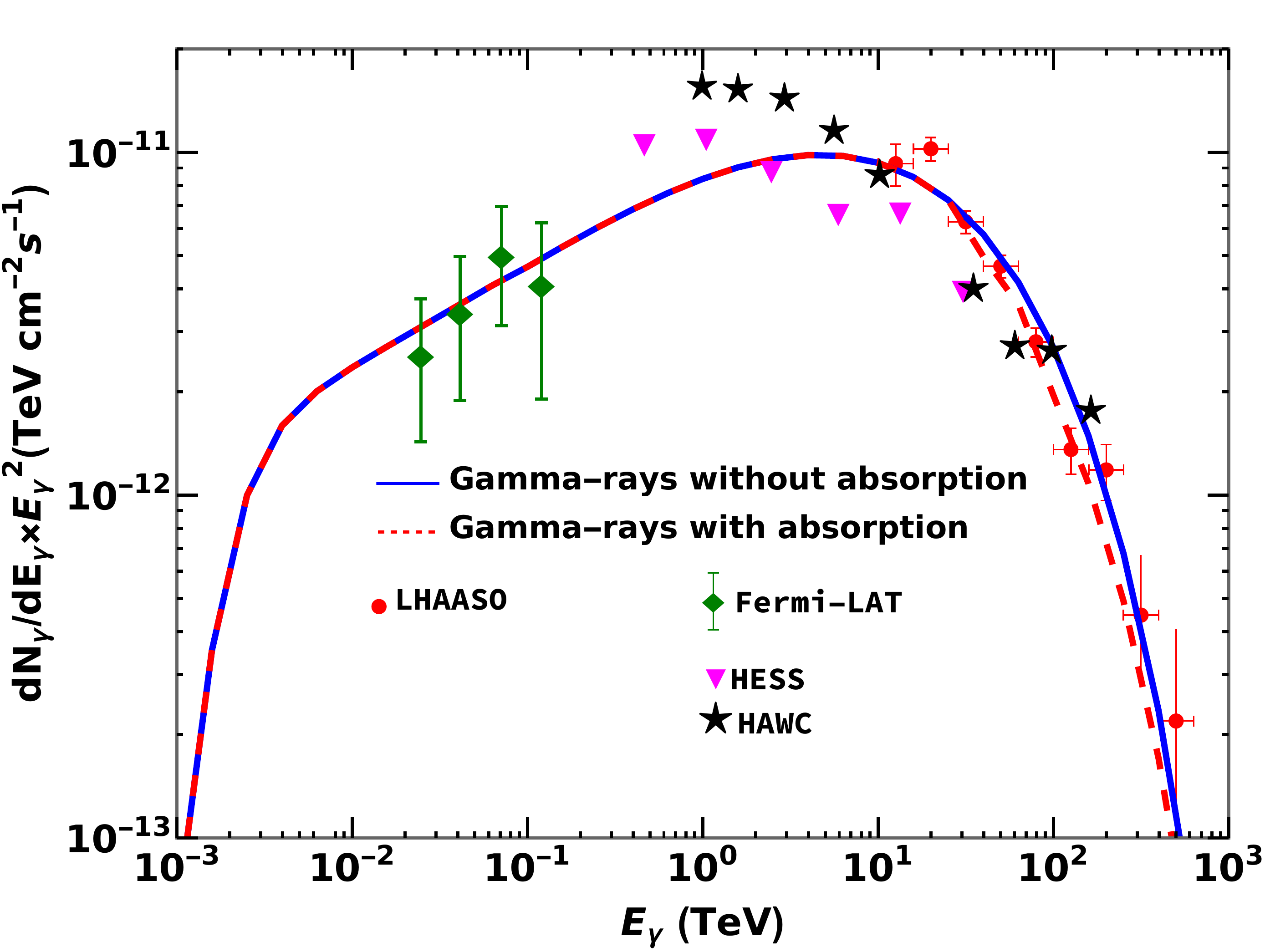}~
 \includegraphics[width=0.48\textwidth]{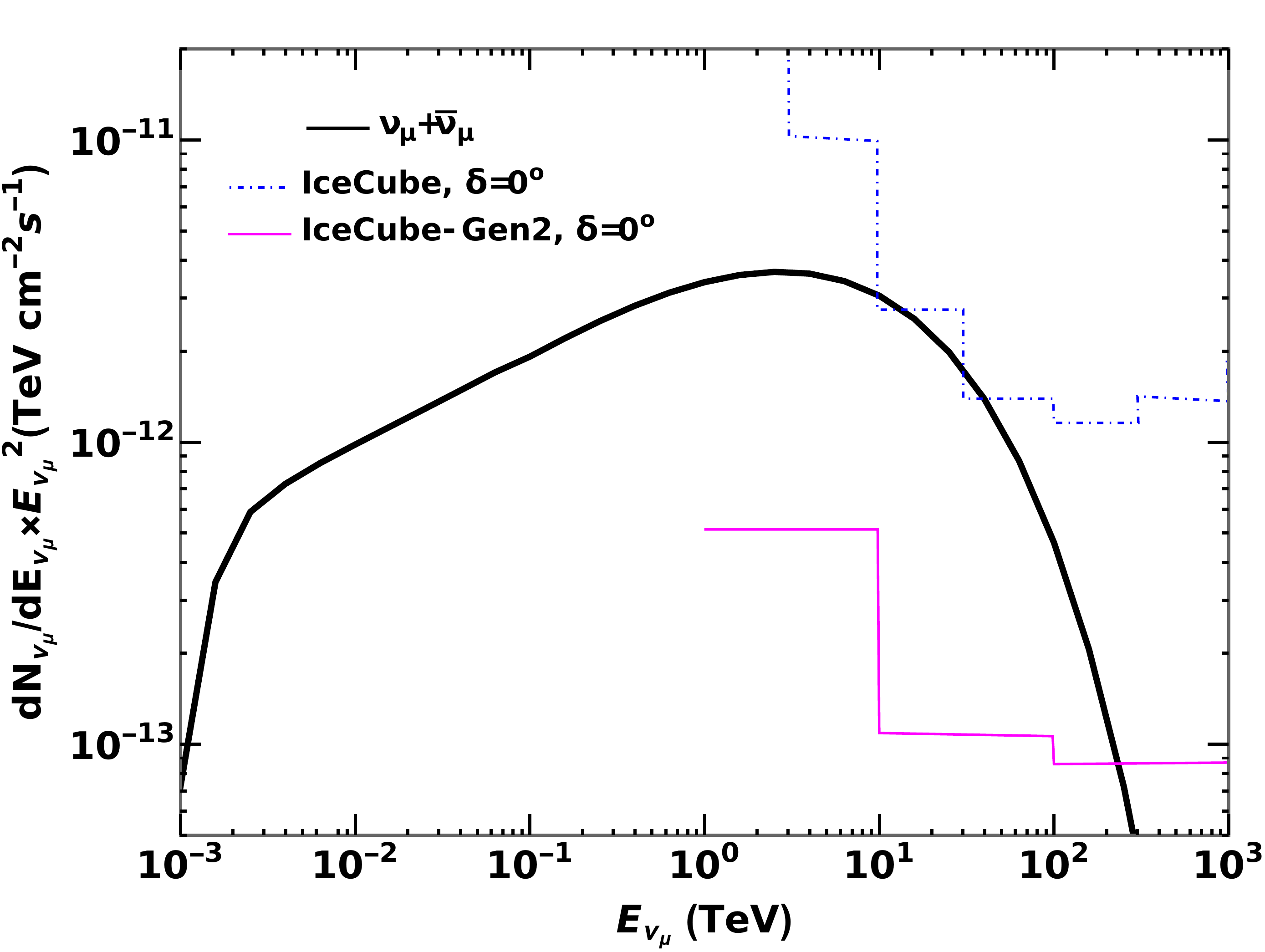}
 }
    \caption{Hadronic gamma-rays (left), and muon neutrinos (right) from SNR G40.5-0.5.  The gamma-ray flux is fitted to the LHAASO J1908+0621 data for a primary proton spectra with index 1.75 and the total energy of protons is $9\times 10^{49}$ erg. The distance to the source is taken to be 8.5 kpc. Gamma-rays above 10 TeV suffer tiny loss due to pair production on ISRF that is shown by the red-dashed curve while the  gamma-ray spectra without absorption is shown by the blue curve. The corresponding neutrino flux has good sensitivity to IceCubt-Gen2 but weak to IceCube. The sensitivities of the neutrino detectors are shown for the declination, $\delta=0^{\degree}$ closest available to the declination of this SNR.}
    \label{fig:G40}
\end{figure*}

\section{Secondary Gamma-ray and Neutrino Fluxes}
\label{sec:model_of_secondaries}
The interaction of CR protons with the molecular cloud leads to the production of charged $\pi^{\pm}$ and neutral pions $\pi^0$, respectively. The decay of neutral pions produces gamma-rays ($\gamma$) and charged pions produce neutrino ($\nu$) fluxes. The fluxes of these gamma-rays and neutrinos depend on the CR proton spectra ($J_{p}(E_{p})$), the molecular cloud density ($N_{H}$), and the production spectra of the secondary particles ($\gamma,~\nu$) for a given energy of proton ($E_{p}$) \citep[][]{2006PhRvD4018K}. The flux of secondary particles is expressed as 

 \begin{equation}
\Phi_{i, pp}(E_{i}) = c n_H \int_{\rm E_{i}}^{\infty} \sigma_{pp}(E_p) J_p(E_p) F_{i}\left(\frac{E_{i}}{E_p}, E_p \right) \frac{dE_p}{E_p},
\label{eq:secondary_fluxes}
\end{equation}

\noindent
where, $i = \gamma~ {\rm or} ~ \nu_{f} $ and $f$ stands for  neutrino flavours ($e$, $\mu$).  The CR protons are assumed to follow a power-law type distribution with an exponential cutoff above $E_0$, i.e., $J_p(E_p) = A_p E_p^{-\alpha} {\rm exp}(-E_p/E_0) $ $\rm TeV^{-1}$. The power-law index $\alpha$ is typically chosen to be 2 based on Fermi's diffusive shock acceleration mechanism \citep[][]{gaisser_engel_resconi_2016}.  The normalization constant $A_{p}$ is related to the total energy ($E_{p,\rm total}$) of the CR protons by the relation $\int_{m_p}^{\infty} E_{p}J_{p}(E_{p}) \mathrm{d}E_{p}=E_{p, \rm total}$ and $\sigma_{pp}(E_{p})$ is the energy dependent cross-section for the $p-p$ interaction \citep{kafe2014PhRvD14K}. In particular, the upper limit of this integration is the maximum proton energy, $E_{p,\rm max}$ which depends on the magnetic field of the acceleration zone, confinement time and shock speed \citep[][]{gaisser_engel_resconi_2016,G1062021NatA60T}. In case of SNR, this $E_{p, \rm max}$ is found to be a few PeV \citep[][]{G1062021NatA60T}. In our analysis, we obtain the proton normalization by fitting the observed gamma-rays, and thus the exact value of $E_{p,\rm max}$ is not necessary for our calculation. The cut-off energy $E_o < E_{p, \rm max}$ makes the proton spectra to fall rapidly for energies larger than $E_{o}$. 

Now, these secondary gamma-rays and the neutrinos are produced from the same $p-p$ interaction implying a connection between the final state fluxes of these secondaries. The secondary fluxes are connected by the following relation at  $E_{\gamma} \approx 2 E_{\nu}$ \citep{Ahler2014PhRvD90010A}

\begin{equation}
E_{\gamma} \frac{\mathrm{d} N_{\gamma}(E_{\gamma})}{\mathrm{d} E_{\gamma}} \simeq e^{-d/\lambda_{\gamma \gamma}} \frac{2}{3} \sum_{f} E_{\nu_f} \frac{\mathrm{d} N_{\nu_f}(E_{\nu})}{\mathrm{d} E_{\nu_f} },
\label{eq:nu-gamma-connection}
\end{equation}

where,  $\frac{\mathrm{d} N_{\gamma}(E_{\gamma})}{\mathrm{d} E_{\gamma}} $ and $\frac{\mathrm{d} N_{\nu_f}(E_{\nu})}{\mathrm{d} E_{\nu_f} }$ are the differential fluxes of gamma-rays and neutrinos of flavour $f$ respectively. $\lambda_{\gamma \gamma}$ is the mean free path for the UHE gamma-rays in the Galactic radiation fields for a source located at a distance $d$. Therefore, one might  predict the neutrino flux from SNR using this multi-messenger connection, i.e., from gamma-ray observations. 

The same secondary gamma-ray and neutrino flux at earth can also be estimated directly from Eq.~\ref{eq:secondary_fluxes} and given by

\begin{equation}
  E_{i}^2  \frac{\mathrm{d}N_{i}}{\mathrm{d} E_{i}} = \frac{E_{i}^{ 2}  \Phi_{i,pp}(E_{i}^{})}{4 \pi d^2} .
  \label{eq:flux_at_earth}
\end{equation}

Since we are interested in detection of muon tracks with  high energy neutrino detectors like IceCube, the muon flux at earth, $E_{\nu_{\mu}}^2  \frac{\mathrm{d}N_{\nu_{\mu}}}{\mathrm{d} E_{\nu_{\mu}}}$  is computed by assuming the flavour ratio $1:1:1$. Note that neutrinos and anti-neutrinos are treated as same here. Clearly, for sources with known properties like $n_{H}$, $E_{0}$ etc., one may use Eq.~\ref{eq:flux_at_earth},  otherwise the Eq.~\ref{eq:nu-gamma-connection} can be used for sources with only gamma-ray observations.

Neutrinos being weakly interacting in nature can propagate to earth without any losses. However, gamma-rays may undergo absorption during propagation. Absorption might effect the gamma-rays produced in SNRs and therefore, it is important to analyse this absorption process. This might also effect the cut-off energy, $E_{0}$  of the CR protons i.e., large absorption would allow  large $E_{0}$ for the primary CR flux.  The absorption is due to pair production on low energy photon backgrounds like cosmic microwave background (CMB) and the Galactic interstellar radiation field (ISRF). This is estimated by the factor $e^{-\tau_{\gamma\gamma}}$, where $\tau_{\gamma\gamma}= d/{\lambda_{\gamma\gamma}}$ is the attenuation of gamma-rays on the low energy photons \citep[][]{Moskalenko:2005ng}.   Larger the optical depth, larger is the absorption. The optical depth depends on the density and average energies of these low energy photons and the distance to the source $d$. Clearly, gamma-ray fluxes from further sources experience larger absorption.  The CMB photons' number density is about $440~\rm cm^{-3}$ and has an average energy of the order of $10^{-4}$ eV. This results in threshold energy for gamma-rays to interact with CMB photons at about 100 GeV \citep[][]{Moskalenko:2005ng}. Since the CMB is uniform, the amount of absorption on CMB effectively depends only  on the distance, $d$.  On the other hand, the energies of Galactic ISRF photons are larger than that of CMB photons and the distribution of ISRF photons is highly non-uniform. Therefore, the threshold energy for ISRF is lower than 100 GeV and the absorption depends on the location of the source \citep[][]{Moskalenko:2005ng}. For example, for a source located in a denser ISRF region will have larger absorption than that of a source located in a thinner ISRF. 

In the following, we use  the method of direct calculation, i.e., Eqs.~\ref{eq:secondary_fluxes} and \ref{eq:flux_at_earth} for the secondary fluxes from the Galactic  SNR sources, G106.3+2.7 and G40.5-0.5, as requisite properties are known from observations.

\subsection*{SNR G106.3+2.7}
SNR G106.3+2.7  is one of the probable PeVatron source in which CR protons are accelerated due to the high shock speed \citep{2021Inno200118G}. The distance of this source is 0.8 kpc and its shape is not perfectly spherical but a comet shaped SNR of length 14 pc and width of 6 pc \citep{2001ApJ6036K}. Recently, The Tibet AS$\gamma$ Collaboration reported the discovery of gamma-rays above 10 TeV which extends upto 100 TeV gamma-rays \citep{G1062021NatA60T}. Further, the LHAASO collaboration also reported the discovery of UHE gamma-rays from this spatial region from the source LHAASO J2226+6057 \citep{cao2021Natur33C}.  These UHE gamma-rays are found to be spatially correlated with a molecular cloud and hence CR proton interactions are considered to be strong contenders \citep{G1062021NatA60T}. This is also supported by the hadronic interaction models developed for this source \citep{2021ApJ32B, 2022AnA60Y}.  The probable density of the molecular cloud region is approximately $10~ {\rm cm^{-3}}$  and  is consistent with the Bremsstrahlung radiation of electrons \citep{2021ApJ1133F}.

Figure \ref{fig:G106} shows the model fitting of Tibet AS+MD gamma-ray data (left) and the corresponding neutrino emission in the model (right). The multi-messenger nature of the emission is crucial in fitting the gamma-ray spectra. The spectra data at the lower energies ($10^{-3}-1$ TeV) are from VERITAS \citep{2009ApJ...703L...6A}, at intermediate energies ($1-10$ TeV) are from Fermi-LAT \citep{2019ApJ...885..162X}, and at higher energies ($\sim 10$ TeV) are taken from Tibet AS$\gamma$ \citep{G1062021NatA60T}. The best fitting parameters $E_{p, \rm total}$, $E_0$ and $\alpha$ are such that both the normalisation and the slope of the spectra are well explained (see Table~\ref{tab:parameters}). We note the smaller $E_{p, \rm total}$ for this SNR and this may be due to low  transfer of shock kinetic energy to accelerated protons \citep[][]{G1062021NatA60T}. Moreover, the power law index, $\alpha$ in this object is found to be less than 2 from gamma-ray spectra fitting. This is in contrast with the diffusive shock acceleration theory that predicts $\alpha = 2$. Extreme pressure at shock front might cause such efficient particle acceleration \citep[see e.g.,][]{1999ApJ...511L..53M}. These fit parameters are also consistent with \citep{G1062021NatA60T}. The source being located only at 0.8 kpc,  the gamma-ray flux has negligible absorption during propagation to earth \citep[][]{G1062021NatA60T,Moskalenko:2005ng}.

Based on these parameter choices, we estimate the gamma-ray flux from this object. The resulting gamma-ray flux gives a good fitting to the gamma-ray data at both low energy and high energy. As mentioned, absorption during propagation is negligible which resulted in the overlap of the blue and red dashed curves. The associated neutrino flux is estimated using Eq.~\ref{eq:flux_at_earth} and is found to be consistent with Eq.~\ref{eq:nu-gamma-connection}. This neutrino flux is found to be much below the detection sensitivity of IceCube. However, IceCube-Gen2 will be able to detect these neutrinos provided the source is located closer to earth.   Note that larger $\alpha$ would change the sensitivity.

\subsection*{SNR G40.5-0.5}

SNR G40.5-0.5 is spatially overlapped with the UHE gamma-ray source LHAASO J1908+0621 \citep{cao2021Natur33C}. In earlier observations, this SNR was found be overlapped with extended 
TeV source MGRO J1908+06,  detected by Milagro Galactic plane survey  \citep{2007ApJ664L91A}. This source was also detected by the H.E.S.S. telescope in gamma-rays \citep{2009AnA723A}. UHE gamma-rays were also discovered from MGRO J1908+06 source by the HAWC detector \citep{PhysRevLett.124.021102} that makes SNR G40.5-0.5 a plausible CR particle accelerator. The source distance is uncertain and CO Molecular-line emission infers a source distance of approximately 3.4 kpc \citep{2006ChJAA210Y} but at a larger distance of 5.5 to 8.5 kpc using $\Sigma-$D relation \citep{downes1980AnA47D}. The mean gas density in the associated molecular cloud is $\sim 45 ~{\rm cm^{-3}}$  taking the source distance 8 kpc \citep{2021ApJ33L} and the source linear size at this distance would be approximately 60 pc.

Model fit of the gamma-ray data of this object is shown in the left panel Fig.~\ref{fig:G40} for the parameters  listed in Table~\ref{tab:parameters}. The low energy ($10^{-3}-10^{-1}$ TeV) data are from Fermi-LAT \citep{2020ApJS..247...33A,cao2021Natur33C} and at high energies ($>10^{-1}$ TeV) data are taken from HESS \citep{2018A&A...612A...1H}, HAWC \citep{PhysRevLett.124.021102}, and LHAASO \citep{cao2021Natur33C}. The   parameters   listed in Table~\ref{tab:parameters}  can well explain these multi-messenger gamma-ray data. Note that  $E_{p, \rm total}$ in this SNR is found to be larger than that of SNR G106.3+2.7. However, the power-law index $\alpha$ is found to be similar to that of G106.3+2.7. Since this source is farther away,  gamma-rays suffer larger absorption due to pair production on ISRF compared to G106.3+2.7. However, this absorption is still tiny and negligible; and shown by the red dashed curve. The corresponding neutrino flux is estimated using Eq.~\ref{eq:flux_at_earth} and shown in the right panel. This neutrino flux estimation is found to be consistent with Eq.~\ref{eq:nu-gamma-connection}. To check the IceCube neutrino sensitivity, we look into the closest available declination of IceCube, i.e., $\delta=0^{\degree}$  to the source declination.  The neutrino flux has weak sensitivity to IceCube but has reasonably good sensitivity to IceCube-Gen2. Hence, this object sets an outstanding goal to neutrino detection in IceCube-Gen2 as the neutrino flux is much above the detection sensitivity. 

In the following, we investigate the parameter space of Galactic PeVatrons allowed by gamma-ray observations \citep[][]{G1062021NatA60T,cao2021Natur33C} and discuss the future detection prospects with gamma-ray and neutrino detectors.

\begin{table}
\centering
\begin{tabular}{|c|c|c|c|}
\hline
{\bf Parameters} & {\bf SNR G106.3+2.7} & {\bf SNR G40.5-0.5} \\
\hline
$E_{p, \rm total}$ (erg) & $8\times 10^{47}$ & $9\times 10^{49}$\\
\hline
$E_{_0}$ (TeV) & $6\times 10^2$ & $4\times 10^2$ \\
\hline
$\alpha$ & 1.8 & 1.75 \\
\hline
\end{tabular}

 \caption{Model parameters chosen for the two prototype SNRs.  These parameters give the best fit to the observed gamma-ray data of these objects. SNR G106.3+2.7 is located at 0.8 kpc and G40.5-0.5 is located at 8.5 kpc. The target gas density, $n_{H}$ for SNR G106.3+2.7 is $10~ \rm cm^{-3}$ and for G40.5-0.5 is $45~\rm cm^{-3}$.}
 \label{tab:parameters}
\end{table}

\begin{figure*}
    \centering
     \includegraphics[width=0.48\textwidth]{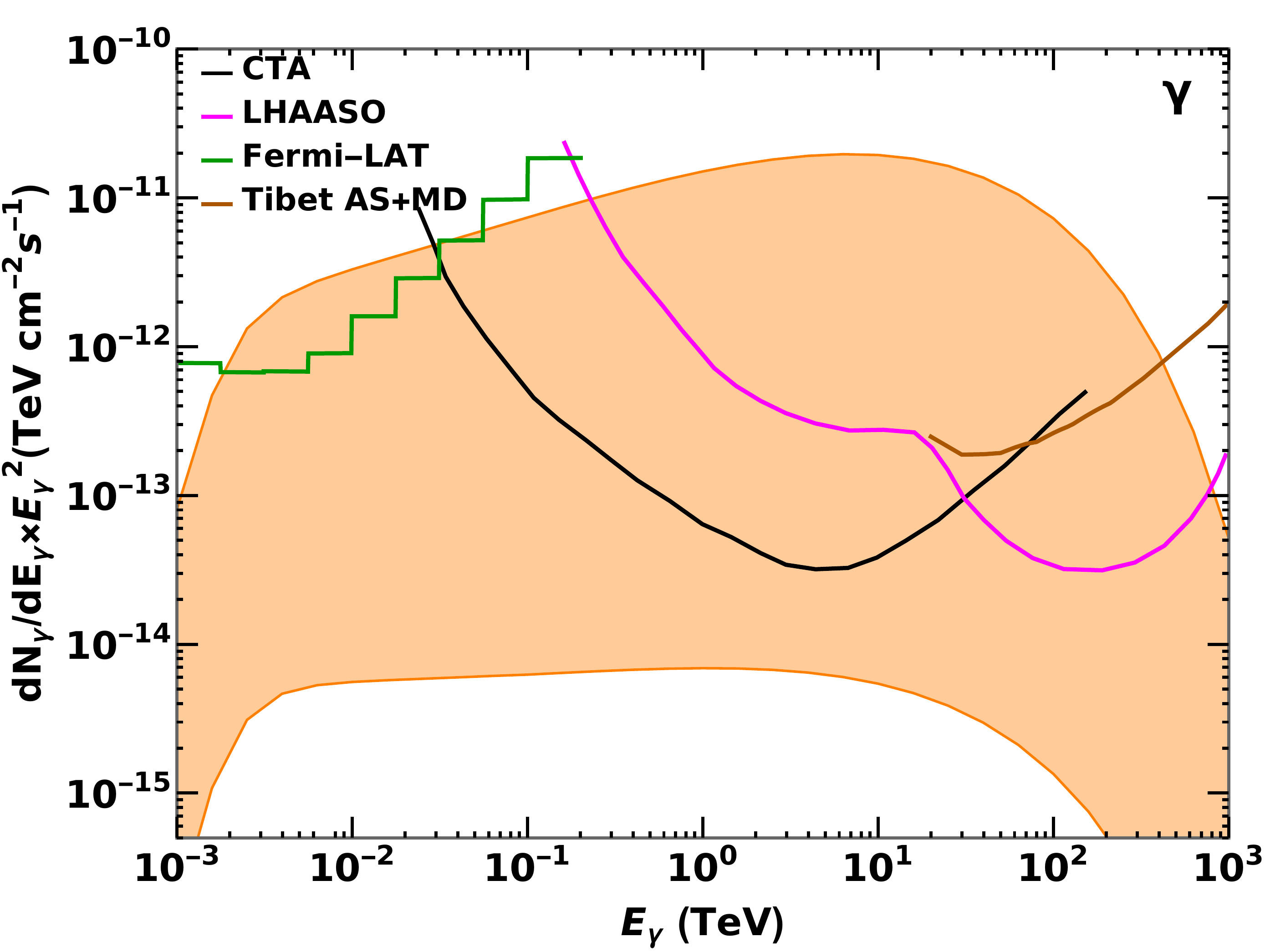}~
    \includegraphics[width=0.48\textwidth]{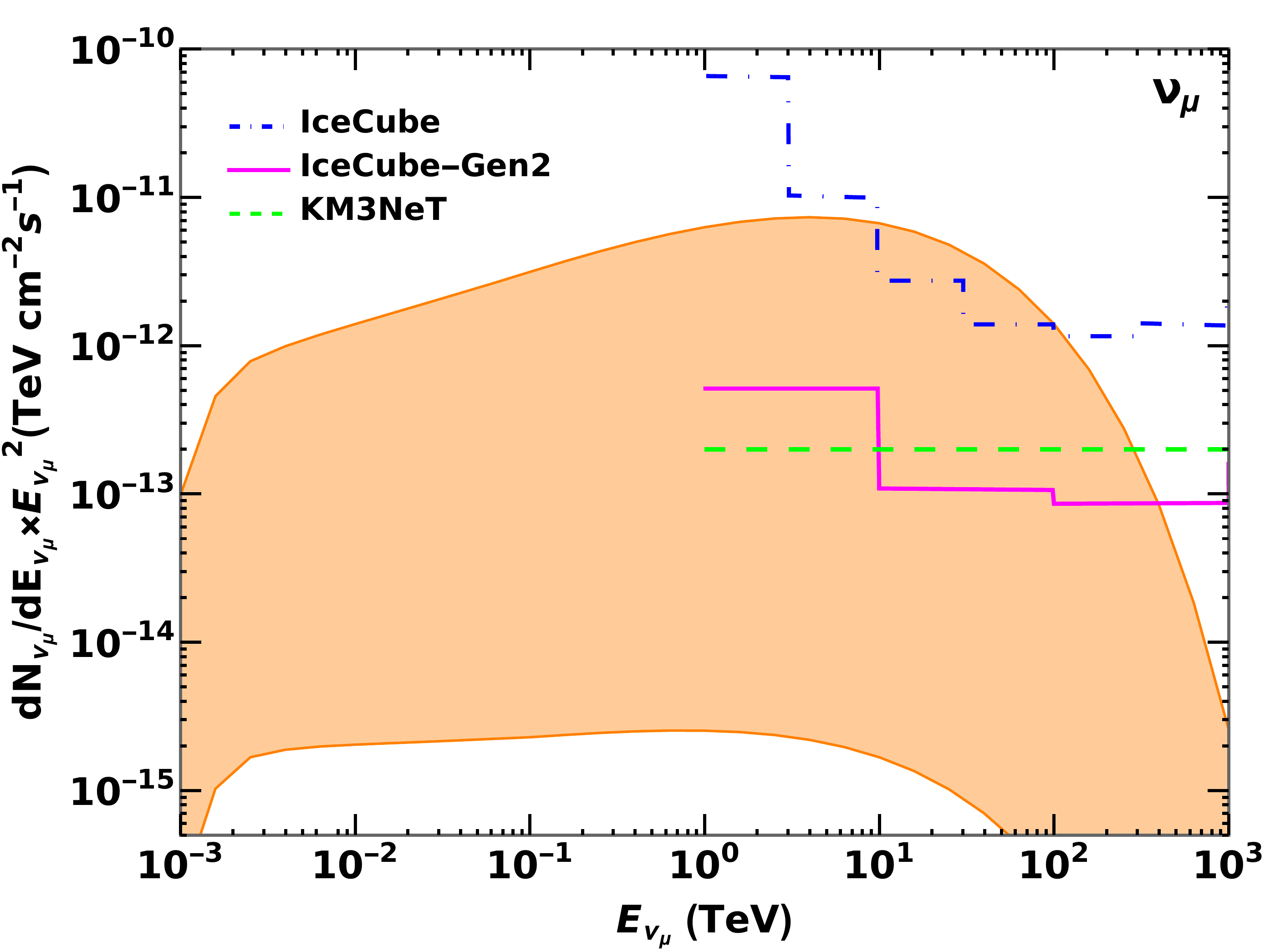}
    \caption{{\it Left:} Detection prospects of gamma-rays from Galactic SNR at 10 kpc with different gamma-ray telescopes. The curves in different colours shows the sensitivities of Fermi-LAT (green) \citep[][]{Fermi-LAT_Sensitivity}, LHAASO (magenta) \citep[][]{Vernetto_2016}, Tibet AS+MD (brown) \citep[][]{Tibet-ASG} and CTA (black) \citep{CTA}.  The orange band shows gamma-ray flux for the parameter space based on the observations of SNR G106.3+2.7 and G40.5-0.5. This shows that the high energy gamma-ray  detectors will probe large fraction of this parameter space for SNR at 10 kpc. {\it Right:} Corresponding detection prospects at high energy neutrino detectors. The sensitivities of IceCube, IceCube-Gen2 and KM3NeT are shown by the blue dot-dashed, magenta and green lines respectively \citep[][]{2021arXiv210109836I,IceCube-Gen2:2020qha,KM3NeT:2018wnd}. These neutrino detectors will also probe some part of the parameter space.  
    }
    \label{fig:detectionGalacticSNR}
\end{figure*}

\begin{figure*}
    \centering
        \includegraphics[width=0.48\textwidth]{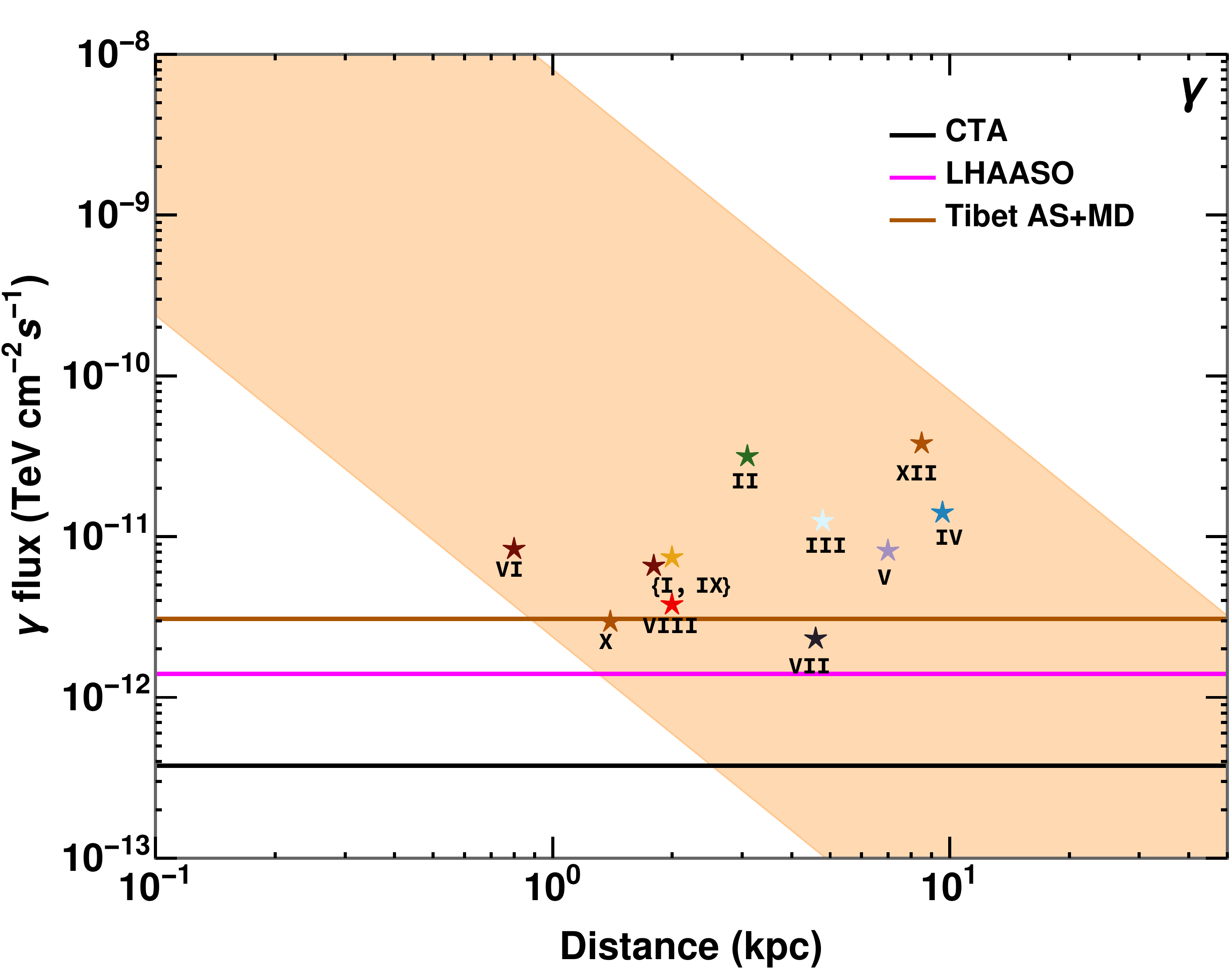}~
    \includegraphics[width=0.48\textwidth]{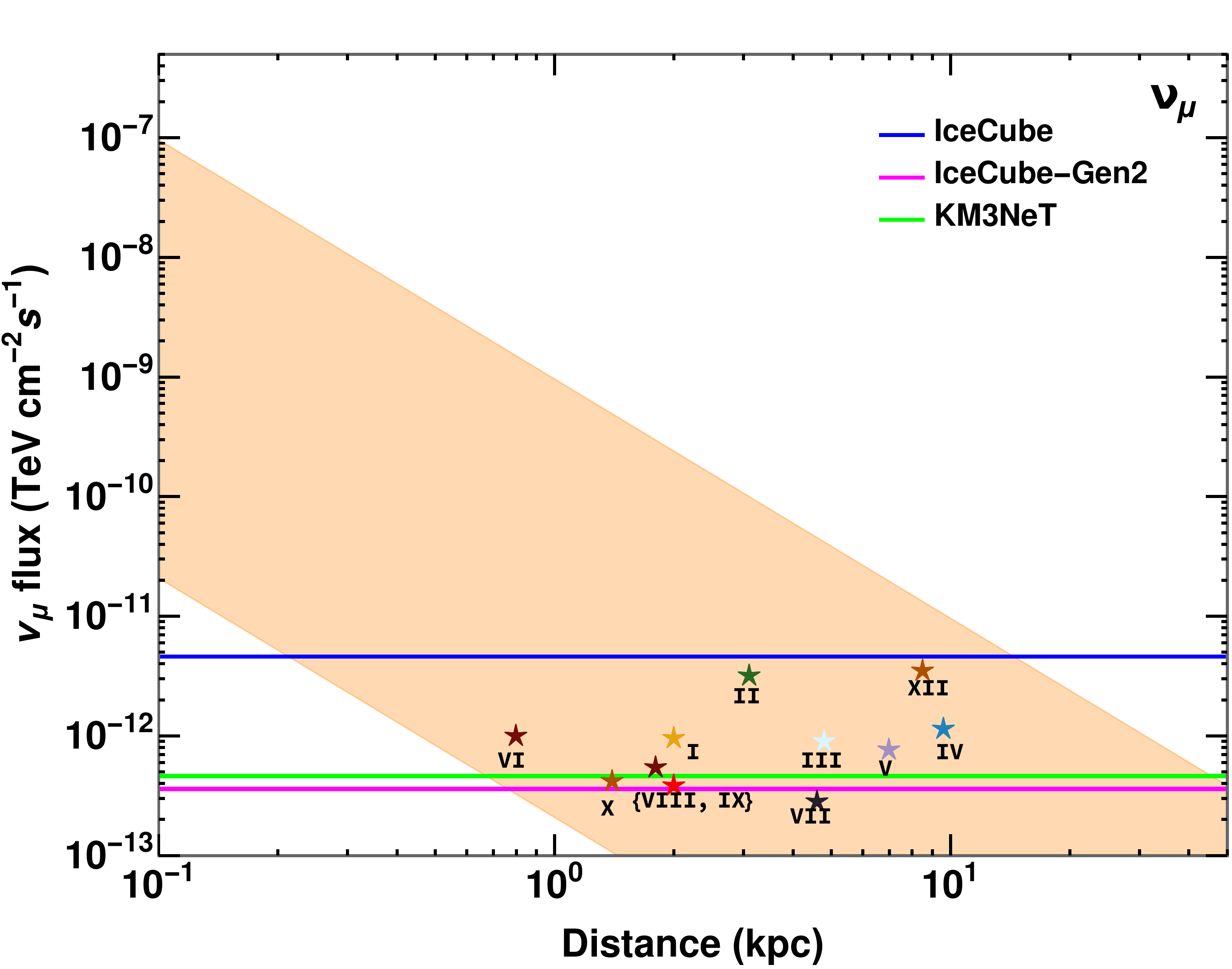}
    \caption{ {\it Left:}  Detection horizon of SNRs for different gamma-ray telescopes. The orange band show integrated flux above 1 TeV as a function of distance (kpc) for the same parameter as in Fig.~\ref{fig:detectionGalacticSNR}.. The horizontal lines show the detectors sensitivities \citep{Vernetto_2016,Fermi-LAT_Sensitivity,CTA,Tibet-ASG}.  All these gamma-ray telescopes are capable of probing a large part of the parameter space . The upcoming CTA will probe sources as far as 50 kpc.  {\it Right:} Corresponding detection horizon of different neutrino detectors. The orange band shows the energy integrated ($10^1-10^2$ TeV) $\nu_{\mu}$ flux as a function of distance (kpc) for the same parameter as in Fig.~\ref{fig:detectionGalacticSNR}. The horizontal lines show the energy integrated sensitivities (corresponds to $\delta=0^{\degree}$ in the energy range $10^1-10^2$ TeV) of IceCube (blue), IceCube-Gen2 (magenta)  and KM3NeT (green) \citep[][]{2021arXiv210109836I,IceCube-Gen2:2020qha,KM3NeT:2018wnd}. This shows that the maximum reach of IceCube  is about 10 kpc whereas for IceCube-Gen2 and KM3NeT, the maximum reach is about 50kpc. The stars of different colours in both panels show the  energy integrated $\gamma$ and $\nu_{\mu}$ flux for the sources listed in Table~\ref{tab:LHAASO}.
    }
    \label{fig:detectionHorizon}
\end{figure*}

\section{Detection prospects of Galactic PeVatrons}
\label{sec:detection}
In the preceding sections, we have computed the expected muon neutrino flux in detail for the two prototype sources. The neutrino flux estimates of such objects show that the flux  might be detected by the current and upcoming generation of IceCube experiment.  This motivates one to look further into the detection prospects of these Galactic objects.  Based on this,  we make  estimate of  future detection of such sources motivated by the two detected prototype SNRs.  We also estimate  the expected neutrino events in IceCube  from the  LHAASO catalog sources (see Table~\ref{tab:LHAASO} and \citep[][]{cao2021Natur33C}) to check the consistency of our chosen parameter space. 


In order to investigate the detection possibility, a range of CR proton spectra parameters is considered, motivated by the observations \citep[][]{cao2021Natur33C,G1062021NatA60T}. This parameter space is obtained by varying the total energy $E_{p, \rm total}$ ($10^{48}-10^{51}$ erg) and the power law index $\alpha$ ($1.7-2.0$) based on the gamma-ray observations of SNR G106.3+2.7 and G40.5-0.5. While varying the parameters, we choose to keep the  cut-off energy, $E_{0}$ fixed as both objects show similar values. The power law index, $\alpha$ is also found to be similar in these two objects, i.e., about 1.8. However, this index is in contrast with the diffusive shock acceleration theory which predicts a power law index $\alpha=2.0$ \citep[][]{gaisser_engel_resconi_2016}. Such smaller value of $\alpha$ is possibly due to very efficient proton acceleration in these objects \citep[][]{1999ApJ...511L..53M,2002APh429B}.   Indeed, observation of neutrinos from such sources by the neutrino detectors in question will be crucial to probe the softer spectra i.e., large $\alpha$. Therefore, the variation of $\alpha$ is considered in the range $(1.7-2.0)$.

For the parameter space discussed above, gamma-ray and muon neutrino fluxes are estimated for an SNR at 10kpc. The left panel in  Fig.~\ref{fig:detectionGalacticSNR} shows the gamma-ray flux and the right panel shows the corresponding muon neutrino flux. The orange band in both panels corresponds to the uncertainty in the parameters $E_{p, \rm total}$ and $\alpha$. In the gamma-ray plot, the sensitivities of different gamma-ray telescopes operating at  different energy ranges have been shown. At low energy, Fermi-LAT is capable of probing a small part of the SNR parameter space. Interestingly, the high energy gamma-ray telescopes like LHAASO, Tibet AS+MD and CTA (upcoming) will be able to probe a much larger fraction of the SNR parameter space.

In addition to the correlated neutrino flux, high energy neutrino detectors'  sensitivities  have also been plotted in right panel of Fig.~\ref{fig:detectionGalacticSNR}. The blue dot dashed line corresponds to the sensitivity of IceCube \citep[][]{2021arXiv210109836I}, the magenta line shows the sensitivity of  IceCube-Gen2 \citep[][]{IceCube-Gen2:2020qha} and the green-dashed line is the sensitivity of KM3NeT \citep[][]{KM3NeT:2018wnd}.  This shows that IceCube is hardly sensitive to such neutrinos from sources at 10 kpc. However, IceCube-Gen2 and KM3NeT might be able to detect these neutrinos from SNRs provided the neutrino fluxes produced are  $> (10^{-13}) \rm ~ TeVcm^{-2}s^{-1}$ above 10 TeV.  Nevertheless, these future detectors together with the gamma-ray telescopes can put stringent constraints on a significant region of the SNR parameter space. For this, the UHE gamma-rays detected by LHAASO \citep[][]{cao2021Natur33C} might assist to scan this parameter space and help us to understand the future detection possibilities.

\begin{table}
\centering
\begin{tabular}{|c|c|c|c|}
\hline
{\bf Sl No.} & {\bf LHAASO sources} & {\bf  IceCube events, $\mathcal{N}_{\nu_{\mu}}$ (5 yr)} \\
\hline 
\Romannum{1} & J0534+2202 & 2.74\\
\hline
\Romannum{2} & J1825-1326 & 0.28 \\
\hline
\Romannum{3} & J1839-0545 & 0.08 \\
\hline
\Romannum{4} & J1843-0338 & 4.68 \\
\hline
\Romannum{5} & J1849-0003 & 2.82 \\ 
\hline
\Romannum{6} & \bf J1908+0621 & 4.55\\
\hline
\Romannum{7} & J1929+1745 & 0.85\\
\hline
\Romannum{8} & J1956+2845 & 1.33 \\
\hline
\Romannum{9} & J2018+3651 & 2.33 \\
\hline
\Romannum{10} & J2108+5157 &  1.05\\
\hline
\Romannum{11} & J2032+4102 & 1.07 \\
\hline
\Romannum{12} & \bf J2226+6057 & 2.57\\
\hline
\end{tabular}

 \caption{Expected number $\nu_{\mu}$ events at IceCube for the LHAASO catalogue. The sources J2226+6057 and J1908+0621 are associated with SNR G106.3+2.7 and G40.5-0.5 respectively. The conventional atmospheric $\nu_{\mu}$ flux produces a background of $\mathcal{O}(10^4)$ events on top of these events.}
 \label{tab:LHAASO}
\end{table}

The PeVatron sources detected in gamma-rays by LHAASO \citep[][]{cao2021Natur33C} are listed in Table~\ref{tab:LHAASO}. The estimation of neutrinos from these sources is not straightforward due to the following reasons. All of these objects might not be pure hadronic PeVatrons  and therefore might not produce much neutrinos \citep[][]{2022arXiv220903970S}. In fact, the source J0534+2202 is found to be associated with the Crab nebula and most of these gamma-rays are expected to be of leptonic origin except for the tail of the spectra which might be hadronic \citep[][]{2021Sci37425L}. Therefore, we use the gamma-ray-neutrino correlation given by Eq.~\ref{eq:nu-gamma-connection}  to estimate the maximum possible neutrinos from the  Crab considering these gamma-rays to be completely hadronic. For the remaining LHAASO sources,  association of any molecular cloud is not well known \citep[][]{cao2021Natur33C} and model of CR interacting with molecular cloud might not be appropriate. In addition, only the fluxes at 100 TeV are available for these sources  \citep[][]{2021Sci37425L}. Thus, we  assume a proton power law index $\alpha=1.8$ and  normalize the gamma-ray flux at 100 TeV to estimate the neutrino fluxes from these objects \citep[][]{cao2021Natur33C}.

To analyze  the detection possibility of these sources with IceCube, we have computed the number of $\nu_{\mu}$ tracks at IceCube by the following formula

\begin{equation}
    \mathcal{N}_{\nu_{\mu}} = T \int_{0.1~TeV}^{100~ TeV} \mathrm{d}E_{\nu_{\mu}} \frac{\mathrm{d} N_{\nu_{\mu}}}{\mathrm{d}E_{\nu_{\mu}}} A_{eff} (\delta) \,
\end{equation}
where, $T$ is the observation time  and $A_{eff}(\delta)$ is the declination dependent effective area \citep[][]{Aartsen_2017}.  The declination of the source plays an important role in the detection of neutrinos. Therefore, the declination of each source in Table~\ref{tab:LHAASO} is taken into account for the calculation of IceCube events. The number of events for 5 years are  shown  in Table~\ref{tab:LHAASO}. The events are found to be so tiny that they might be buried deep in the large conventional atmospheric background ($\mathcal{O}(10^4)$). This explains the non-observation of neutrinos till date from such SNRs \citep[][]{2022arXiv221114184A}.

It is clear from the above analysis that the source distance is also an important parameter for the detection of these PeVatron  sources by gamma-ray and neutrino telescopes.  Fig.~\ref{fig:detectionHorizon} shows the energy integrated gamma-ray (left) and muon neutrino (right) flux as a function of source distance (in kpc). The orange band in both panels corresponds to the uncertainty in the SNR parameters,  $E_{p, \rm total}$ and $\alpha$ as discussed above.  The gamma-ray flux is integrated above 1 TeV following the best sensitive energy ranges of the different detectors. The horizontal lines in the left plot are the integral sensitivities for CTA, LHAASO and Tibet AS+MD. All these gamma-ray telescopes are well capable of detecting SNRs in the galaxy. In future, CTA will probe vast region of the SNR parameter space.

Similarly, detection horizons of different high energy neutrino telescopes like IceCube, IceCube-Gen2 and KM3NeT are shown in the right panel of  Fig.~\ref{fig:detectionHorizon}. The orange band shows  the integrated $\nu_{\mu}$ flux in the range ($10-10^2$) TeV  for the parameter space discussed above. Energy integrated sensitivities in the range $10^1-10^2$ TeV of IceCube, IceCube-Gen2 and KM3NeT for declination angle $\delta=0^{\degree}$ have also been plotted. 
This plot shows that the maximum reach of IceCube  is about 10 kpc whereas for IceCube-Gen2 and KM3NeT the maximum reach is about 50kpc.  However, for the lower limit of the neutrino flux, the detection horizon for IceCube falls below 200 pc and for IceCube-Gen2, KM3NeT, it falls below 1 kpc.

The gamma-ray and neutrino fluxes produced by the LHAASO sources \citep[][]{cao2021Natur33C} listed in Table.~\ref{tab:LHAASO} are also shown in Fig.~\ref{fig:detectionHorizon} by the star symbols in different colours. All these sources are above the sensitivities of LHAASO and CTA, while they are below the sensitivity of IceCube.   Interestingly, the upcoming neutrino detectors IceCube-Gen2 and KM3NeT will be able to detect neutrinos from such SNRs. As pointed out earlier, all of these sources might not be pure hadronic PeVatrons \citep[][]{2022arXiv220903970S}.   Therefore, the neutrino fluxes shown here can only be treated  as upper limits.  As mentioned earlier, the Crab gamma-ray fluxes are expected to be mostly leptonic \citep[][]{2021Sci37425L}. Thus, neutrino emission from Crab might be faint and not detectable.  Even for the purely hadronic model, the neutrino flux from Crab is found to be below IceCube sensitivity. Indeed, IceCube-Gen2 and KM3NeT will be able to constrain the hadronic nature of these sources.

Note that the detection of neutrinos strongly depends on the source declination, $\delta$. The sensitivities shown here for the neutrino detectors are for $\delta=0^{\degree}$, i.e., the largest sensitivities. For declination angle other than this, the sensitivity would be lower and some of the sources might fall below the detectors' sensitivities. Nevertheless, the upcoming neutrino detectors might still be able to detect some PeVatrons and unravel the nature of these sources, i.e, hadronic or leptonic.

\section{Discussions and Conclusions}
\label{sec:conclusion}
The 12 sources detected by LHAASO in TeV gamma-rays provide evidence of CR acceleration to very high energies. If these gamma-rays are created by hadronic interaction of CR protons with background protons, they should be accompanied by neutrinos. However, no neutrinos are expected for leptonic origin of these gamma-rays. Therefore, detection of neutrinos will play a crucial role to identify the nature of these sources. Based on the gamma ray data, we have estimated the plausible neutrino fluxes from these sources and test their detectability in IceCube, IceCube-Gen2 and KM3NeT.

These objects detected by LHAASO are found to be spatially associated with SNRs and pulsars.  CRs accelerated in SNRs and pulsar  interacting with any nearby molecular cloud can be the origin of these gamma-rays. In fact, for two of these sources, gamma-rays are  found to be originated in the molecular clouds of SNR G106.3+2.7 and G40.5-0.5. However, for some of these sources no molecular clouds have been identified yet. On the other hand, one of these objects is found to be connected to the Crab nebula where the emission is expected to be mostly leptonic.  Therefore, detection of the associated neutrino signal is important to probe the nature of these sources. 

Using the gamma-ray data of SNR G106.3+2.7 and SNR G40.5-0.5, we have developed a $p-p$ interaction model and estimated the neutrino fluxes from these objects.  The age of SNR G106.3+2.7 is $\sim 10$ kyr, while for SNR G40.5-0.5, it is in between 10-20 kyr \citep[see e.g.,][]{cao2021Natur33C}.  For these two sources, model parameters, i.e., CR proton energy $E_{\rm p, \rm total}$, spectral index $\alpha$, the cutoff energy $E_0$ are listed in Table \ref{tab:parameters}.  We found $\alpha$ and $E_{0}$ to be similar for both SNRs (1.8 and $\sim 500$ TeV) but their $E_{\rm p, \rm total}$ is very different. The asymmetry in the source geometry and cloud morphology can affect the escape of the CR protons and may cause this difference in the total energy in CR protons \citep{2021ApJ1133F}. Our chosen spectral indices are harder compared to the standard diffusive shock acceleration model, which predicts $\alpha =2$ \citep{1983RPPh973D}. However, if the shock compression ratio is very high then the CR spectral index can be harder; i.e. $\alpha =1.5$ \citep{1999Ap11L53M} as also predicted using nonlinear diffusive shock acceleration mechanism \citep{2002APh429B}. The neutrino flux from SNR G106.3+2.7 is found to be below IceCube sensitivity and SNR G40.5-0.5 has poor sensitivity to IceCube. This also explains the non-observation of neutrinos from these two sources. However, both these sources appeared to be sensitive to IceCube-Gen2. Hence, IceCube-Gen2 will be able to probe the hadronic channel in these objects.

Based on the analysis of these two sources, we have derived the probable parameter space of this SNR + molecular cloud model. The parameter space has been constructed considering uncertainties in the model parameters, i.e., spectral index, $\alpha$ and total energy, $E_{p, \rm total}$. The upper limit of $\alpha=2.0$ is considered based on the standard CR acceleration mechanism. The wide variation in $E_{p,\rm total}$ is taken based on the observation the two SNR G106.3+2.7 and G40.5-0.5.  Future neutrino telescopes (IceCube-Gen2 and KM3NeT) will be able to probe a large portion of this parameter space, provided the events are located within 50 kpc. However, the best possibility of detection indeed lies within a region of 10 kpc which is inferred by the LHAASO sources.

We have also estimated the neutrino flux from Crab nebula based on the  gamma-ray spectra observed by LHAASO. Most of these gamma-rays are expected to be of leptonic origin.  Thus, the neutrino signal is expected to be weak. The neutrino flux from Crab is found to be below IceCube sensitivity, even for the purely hadronic model. Neutrino fluxes for the remaining LHAASO sources are estimated based on the gamma-ray data and assuming a power law index 1.8. All these sources are found to be below IceCube sensitivity but many of them might be detectable in IceCube-Gen2 and KM3NeT.

Further, primary and secondary electrons accelerated at the shock regions, can also contribute towards the multi-wavelength radiation. Their synchrotron radiation depends on the magnetic field and hence this can be tuned to get the radio to X-ray spectrum consistent with the observations. The inverse-Compton radiation can be another channel for the gamma-ray spectrum in the  UHE regime. In this work we are mainly interested in estimating the maximum neutrino flux from SNRs, hence we have not discussed the radiation due to electrons. 
However, contamination from the leptonic channel of the concentric PWN cannot be discarded in these sources \citep{2014JHEAp...1...31T, 2022NewA001669Y,2022arXiv220500521J}

To conclude, we have analysed the detection prospects of the LHAASO sources and also similar possible sources in the Galaxy. IceCube might be able to detect closer objects within 1 kpc but for extremely neutrino bright sources the detection horizon extends up to 10 kpc. The future detectors KM3NeT and IceCube-Gen2 can have  detection sensitivity up to 50 kpc. Such detection will probe the nature of Galactic PeVatrons.


\section*{Acknowledgements}

 S.C. acknowledges the support of the Max Planck India Mobility Grant from the Max Planck Society, supporting the visit and stay at MPP during the project. S.C has also received funding from DST/SERB projects CRG/2021/002961 and MTR/2021/000540.



\bibliographystyle{mnras}
\bibliography{pevatron_uhe} 






\bsp	
\label{lastpage}
\end{document}